# Integrating energy markets: Implications of increasing electricity trade on prices and emissions in the western United States

Steven Dahlke

September 4, 2019

## Abstract
This paper presents empirically-estimated average hourly relationships between regional electricity trade in the western United States (U.S.) and prices, emissions, and generation from 2015 through 2018. Consistent with economic theory, the analysis finds a negative relationship between electricity price in California and regional trade, conditional on local demand. Each 1 gigawatt-hour (GWh) increase in California electricity imports is associated with an average $0.15 per megawatt-hour (MWh) decrease in the California Independent System Operator's (CAISO) wholesale electricity price. There is a net-negative short-term relationship between carbon dioxide emissions in California and electricity imports that is partially offset by positive emissions from exporting neighbors. Specifically, each 1 GWh increase in regional trade is associated with a net 70-ton average decrease in $CO_2$ emissions across the western U.S., conditional on demand levels. The results provide evidence that electricity imports mostly displace natural gas generation on the margin in the California electricity market. A small positive relationship is observed between short-run $SO_2$ and $NO_x$ emissions in neighboring regions and California electricity imports. The magnitude of the $SO_2$ and $NO_x$ results suggest an average increase of 0.1 MWh from neighboring coal plants is associated with a 1 MWh increase in imports to California.

# 1. Introduction

Those working on research and policy in the electricity sector often think about optimal market designs to meet society's energy goals at the lowest cost. To this end, centralized wholesale electricity markets have grown significantly in the United States (U.S.) over the past two decades. Recent examples include the southward expansion of the Midcontinent Independent System Operator market in 2013, and the northward expansion of the Southwest Power Pool market in 2015. California is now deliberating with neighboring states about whether or not to regionalize its centralized market to increase electricity trade with neighboring states.

The economic, legal, and social impacts of regionalizing California's electricity market have recently been studied by various entities to help inform the political debate. For examples, see Chang et al, 2016; Brint et al, 2017; Hogan, 2017; Tarufelli and Gilbert, 2018. However, because regional market discussions in California have been renewed relatively recently, the current academic literature on the topic is still relatively sparse. This analysis offers new insights, including estimates of recent short-term relationships between increased trade and prices, emissions and electricity supply. Looking to recent history as a reasonable guide, these short-term relationships provide empirically-based estimates of near-term impacts of increasing regional trade across the western U.S. through a regional market.

The analysis finds that from 2015 – 2018, a one gigawatt-hour (GWh) increase in California electricity imports is associated with an average $0.15/MWh decrease in the California Independent System Operator (CAISO) wholesale system electricity price. Extrapolation from this short-term estimate suggests approximately $252 million in annual savings for California electricity consumers if imports were doubled from current levels. This number is calculated from short-run conditional estimates and does not account for future market equilibrium adjustments that will occur, including demand response and changes in the capital stock. However, given the limited demand response in wholesale electricity markets and relatively long time required to develop new electricity resources, short-run effects are relevant until market participants are able to respond via investment decisions and retail rate adjustments. The short-run savings estimates are much larger than the likely market implementation costs. For example a similar regional expansion in the northeastern U.S. required a one-time implementation cost of approximately $40 million.

This study also finds decreased carbon dioxide ($CO_2$) emissions in California associated with electricity imports. These emission reductions are partially offset by increased $CO_2$ emissions in neighboring states as fossil generators increase output when California demand increases. On net, each 1 GWh of increased regional trade is associated with a 70 ton reduction in $CO_2$ emissions across the western U.S. Moreover, the statistical models find that increased regional trade is associated with a small net increase in $SO_2$ and $NO_x$ emissions in neighboring states outside of California, suggesting a relatively small portion of California imports is supplied by out of state coal generation. This result reinforces the importance of having strong emissions policies in place that cover the full regional market, such as the U.S. Acid Rain Program for $SO_2$ emissions. Doing so will keep long-run regional emissions at or below the program cap, despite short run increases in certain areas as regional electricity markets expand.

This paper is organized as follows: section 2 provides general background on electricity markets and recent developments in the western U.S. Section 3 presents relevant economic theory that produces a priori hypotheses on the effects of electricity market integration. Section 4 walks through each step of an econometric analysis. Finally, section 5 discusses policy implications, next steps, and concludes. All the datasets and computer code necessary to replicate the analysis are publicly available and are stored in an analytic appendix online at https://osf.io/hcdn2/.

## 2. Background

For centuries, economists have puzzled over how to structure markets to maximize social welfare. Economic philosophy suggests the value of a market comes from its ability to make information available to both parties involved in an exchange. Efficiency increases when trading partners gain access to additional relevant information. The possession of relevant information allows market participants to reduce uncertainty, identify suitable trading partners and properly negotiate contracts (Hayek, 1945). Moreover, the cost to acquire relevant information and negotiate contracts determines the optimal organization of firms within a market (Coase, 1937; Riordan and Williamson, 1985). In this way, centralized electricity markets are expanding across the U.S. because they increase availability of relevant information to market participants by posting prices, standardizing contracts, and eliminating costs associated with negotiating individual bilateral deals. Centralized markets also eliminate export fees charged by transmission companies for transmitting power across market regions (Chang et al., 2016). An important question for the western U.S. debate is whether the marginal benefits from a centralized wholesale market outweigh the marginal costs of transitioning to such a market. While market implementation costs for the western U.S. are difficult to estimate with precision, Mansur and White (2012) note that a similar market expansion in the PJM region in the northeastern U.S. had a one-time implementation cost of $40 million. This study suggests the immediate consumer savings from transitioning to a regional market largely outweigh costs of this magnitude.

In addition to providing timely information for those working on electricity market policy in the western U.S., this paper builds on a broader scholarship of electricity market integration around the world. In the early 1990's, the European Union issued directives stating their explicit goal of an integrated electricity market, similar to what has occurred recently in California. Since then, there have been many studies evaluating the progress and implications of European electricity market integration towards this goal (Jamasb and Pollitt, 2005; Newbery, Strbac, and Viehoff, 2016). Supplementing this is a body of research evaluating market integration among sub-markets within Europe, including Scandinavia (Amundsen and Bergman, 2007; Lundgren, et al., 2008), southeastern Europe (Hooper and Medvedev, 2009), Italy and its neighbors (Creti et al., 2010), and Ireland and its neighbors (Nepal and Jamasb, 2012). Other work has developed economic models to study effects of electricity market integration in other regions of the world, including eastern Asia (Gnansounou and Dong, 2004; Wu, 2013) western Africa (Gnansounou et al., 2007; Pineau, 2008), and across the western hemisphere (Pineau et al., 2004). Some analysis has been done characterizing the extent of integration within the Western U.S., (Woo et al., 1997; De Vany and Walls, 1999), and more recently on the emissions impacts of increasing integration through western U.S. via recent growth in an energy imbalance market (Hogan, 2017; Tarufelli and Gilbert, 2018). The global literature broadly finds price convergence, reduced volatility, and regional market efficiency benefits after integration, while environmental and production impacts from market integration depend on local resource endowments and supply.

Electricity markets today can broadly be categorized in two ways: Centralized auction markets and decentralized bilateral trading. The market structure in the Western United States varies by state. Trades occur over a grid of electric transmission lines called the Western Interconnection. The Western Interconnection is not synchronized with the eastern United States, and electricity flows between these regions are minimal. In the western U.S. outside of California, the majority of electricity companies are privately-owned firms that are state-regulated monopolies in the locations where they sell power. Most trade between companies utilizes decentralized, bilateral contracts. Bilateral contracts are also heavily utilized to facilitate trade in California, however most electricity is then transacted through a centralized auction market operated by an independent non-profit entity called the California Independent System Operator (CAISO). CAISO collects bids and offers from buyers and sellers in California, and centrally schedules electric generation across the state to meet demand. CAISO also calculates and publishes prices designed to reflect the marginal cost of delivering electricity to each location throughout the state at a given point in time.

Studies of other regions with centralized electricity markets have measured economically significant monetary benefits associated with the market. Mansur & White (2012) estimate $163 million in net gains from trade after expanding the centralized PJM market in the northeastern U.S., leading to roughly a doubling in trading efficiency compared to the bilateral market. Work by Chan et al. (2016) suggests efficiency gains from centralized markets in the U.S. have induced behavioral changes among power plant owners that have led to savings in operations expenses by up to 15%. These past successes have prompted energy policy makers to engage in serious discussions about expanding California's centralized market. In October 2015, California Senate Bill 350, the "Clean Energy and Pollution Reduction Act", was signed into law (De León, 2015). Among other things, this bill established the intent of the California legislature to expand CAISO into a multi-state organization. The legislation required CAISO to study the impact of a regional market, including overall benefits to ratepayers, environmental and emissions impacts, and more. The series of consultant studies referenced in Chang et al. (2016) is the market operator's response to this directive.

This CAISO-commissioned study produced simulations that identified $1-$1.5 billion in annual savings to California consumers from a fully-integrated regional market across the western U.S. Almost half of this is due to operational savings from increased lower-cost imports outside of California, equivalent to the short-run consumer savings empirically estimated in this study. The other half is due to long-run savings from electricity producers' increased ability to meet California demand with large-scale renewable plants that can be more cheaply built outside of the state. These estimated savings are realized by California electricity consumers, and CAISO's simulations do not include effects on California electricity producers, nor consumers and producers in western states outside of California. Economic theory suggests that, all else equal, eliminating barriers to trade across a regional market will decrease consumer costs and producer profits in areas that increase imports, while increasing prices, producer profits and consumer costs in areas that increase exports. The increased price in exporting regions is a significant political-economy constraint that can impede market integration, to the extent that local policymakers represent the interests of their electricity consumers. These issues are discussed further in section 3. California is a net importer of electricity during most hours of the year. As a result, economic theory suggests that increased regional trade will reduce California prices, consistent with the empiric results presented in this paper.

# 3. Economic theory

This section develops a basic economic model to illustrate effects of lowering barriers to regional electricity trade on producers and consumers. The approach is adapted from international trade theory, for example see Suranovic (2010). Consider an electricity market with two regions, a net importing region, like California, and net exporting region, like California's neighbors. Electricity production levels at different prices for each region are represented by a set of supply functions indexed by region $i$ and time period $t$, $S_i(p_{it}, A_{it})$, where $p_{it}$ represents the electricity price and $A_{it}$ represents exogenous factors besides price that determine electricity supply. Wholesale electricity markets often use prices measured in dollars per megawatt-hour ($/MWh). In this case, quantity is measured in megawatts cleared and time period $t$ indexes hours. $A_{it}$ includes fuel costs, government subsidies and taxes, system capacity constraints, and other shocks to the generation and delivery network that impact producer costs. Consumer demand in each market is represented by a demand function $D_i(p_{it}, B_{it})$ where $B_{it}$ represents exogenous factors besides price that determine electricity demand, including changing consumer preferences, weather, prices of related energy goods, income levels, and other shocks that impact demand.

For this simplified example, two market regions are assumed so the set of market regions is $I = \{1,2\}$ The model assumes no constraints to the delivery of electricity within each region, so each region has a single market clearing price. Adding transmission constraints would create price divergence within regions, adding complexity to the model while not changing the general conclusions across regions. The model does assume transmission costs between regions. Transmission costs include charges levied by the transmission owner and transaction costs associated with negotiating individual bilateral deals to trade across regions. Let $c_{ij}$ represent the unit cost to transfer electricity from region $i$ to region $j$. The transfer cost $c_{ij}$ causes price to deviate across regions, such that they satisfy the following conditions in equation set (1).

$$p_{1t} + c_{21t} = p_{2t} + c_{12t} \qquad (1)$$

$$c_{12t} c_{21t} = 0$$

$$p_{1t}, p_{2t}, c_{12t}, c_{21t} \geq 0$$

The equality condition in the second line of equation set (1) restricts electricity transfers to be unidirectional in each period. Thus, if region 1 consumers pay a non-zero $c_{21t}$ to purchase electricity from region 2, there will be no transfers in the other direction, and $c_{12t}$ will equal 0. Next, the equilibrium condition in equation (2) represents the quantity traded from region 1 to region 2 ($q_{12t}$), and ensures that electricity transfers across regions are balanced. Positive values of $q_{12t}$ represent transfers of excess electricity from region 1 to region 2, and negative values represent transfers in the opposite direction.

$$q_{12t} = S_1(\cdot) - D_1(\cdot) = D_2(\cdot) - S_2(\cdot) \qquad (2)$$

Finally, a prohibitive trade cost is defined, equal to the value of $c_{ijt}$, that would make the cost of transferring electricity from region $i$ to region $j$ prohibitively expensive such that no trade occurs. Let the prohibitive trade cost $c_{ijt}^{pro}$ be equal to the smallest transfer cost such that no trade occurs, or $q_{ijt} = 0$. This introduces a set of prohibitive trade cost conditions for the two regions defined in equation set (3).

$$c_{12t} \leq c_{12t}^{pro} \tag{3}$$

$$c_{21t} \leq c_{21t}^{pro}$$

In reality, $c_{ijt}$ can be larger than the prohibitive trade cost. However, from a modeling perspective, any value of $c_{ijt}$ greater than $c_{ijt}^{pro}$ will result in no trade, producing the same result as when $c_{ijt} = c_{ijt}^{pro}$. In both cases, trade is prohibitively expensive and all electricity demand will be met by local supply.

Consumer surplus in each region ($CS_{it}$) is the value of consumption minus the amount paid by consumers, while producer surplus ($PS_{it}$) represents the revenue to producers in excess of marginal production cost (McAfee et al., 2009). Consumer surplus can be calculated as the area under the inverse demand curve and above the market clearing price, while producer surplus can be calculated as total revenue minus the area under the inverse supply curve and below the market clearing price (equation set (4)). The sum of consumer and producer surpluses provides an estimate of the social welfare enabled by the market.

$$CS_{it} = \int_0^{q_{it}} D^{-1}(q_{it}, B_{it}) dq_{it} - p_{it} q_{it} \tag{4}$$

$$PS_{it} = p_{it} q_{it} - \int_0^{q_{it}} S^{-1}(q_{it}, A_{it}) dq_{it}$$

A single period representation of this model is presented graphically in Figure 1. Inverse demand functions are represented by downward sloping lines and inverse supply curves are represented by upward sloping lines. Any functional form for demand and supply with global properties of $\frac{\partial D^{-1}}{\partial q} < 0$ and $\frac{\partial S^{-1}}{\partial q} > 0$ will produce the same general results. In this example, region 1 has relatively higher production costs than region 2, creating an incentive to trade electricity from region 2 to region 1. Panel A shows a situation where trading costs are at or above the level prohibiting trade. Panel B models the market after trading costs have been reduced. In this case, $q_1^D - q_1^S = q_2^S - q_2^D$ of electricity is traded from region 2 to 1. In the real western U.S. electricity market, trading costs would be reduced after the establishment of a centralized regional market due to elimination of transmission access fees and administrative costs associated with negotiating individual bilateral deals.

A summary of the welfare changes from reducing the cost to trade across regions is as follows:

- Consumer surplus in importing region 1 increases ($CS_1^* > CS_1$).
- Producer surplus in the importing region decreases ($PS_1^* < PS_1$)
- Consumer surplus in the exporting region decreases ($CS_2^* < CS_2$)
- producer surplus in the exporting region increases ($PS_2^* > PS_2$).

Furthermore, this model predicts that prices between trading regions will converge after trading costs are lowered. This is shown in Figure 1 as the difference in prices after lowering trading costs ($p_1^* - p_2^*$) is smaller than the difference before ($p_1 - p_2$). In reality, price convergence has been empirically shown in several studies on market integration of European electricity markets (Balaguer, 2011; Kalantzsia and Milonas, 2010; Soares & Pereira da Silva, 2008; Zachmann, 2008).

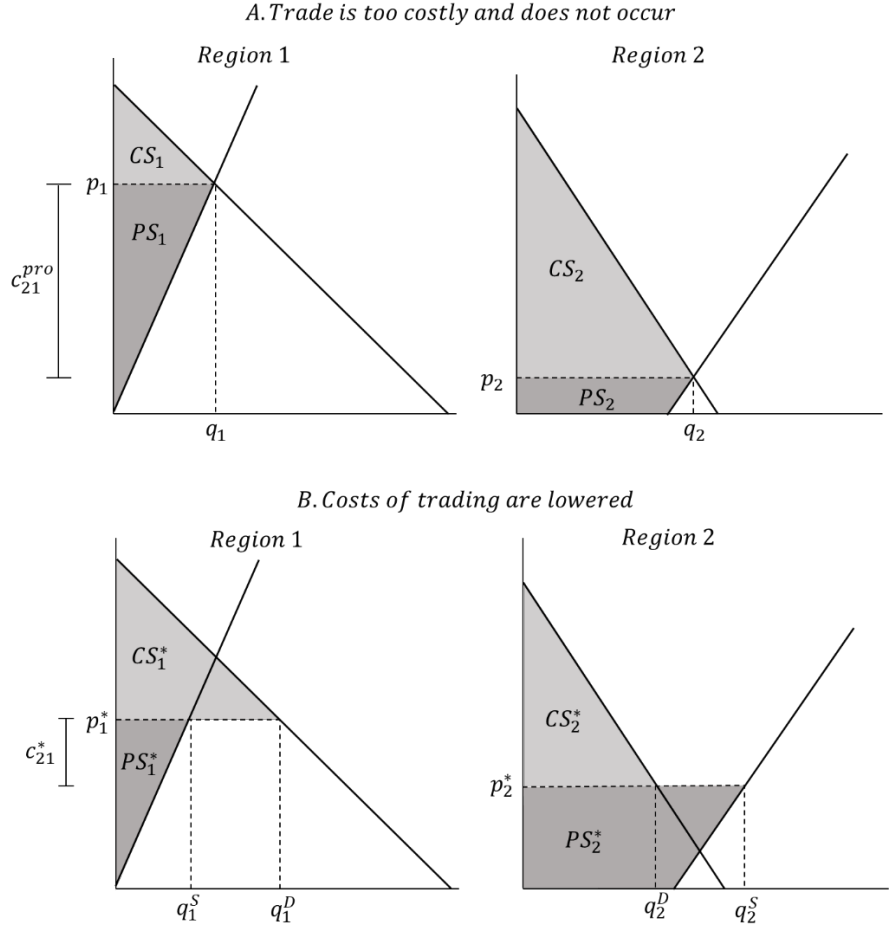

*Figure 1 Effects of a reduction in trading costs on social welfare.*

While California prices are predicted to decrease, neighboring prices with lower-cost electricity generation are predicted to increase. Higher prices would incent investment in new generation capacity, offsetting the short-run price increase. However, in many regions political economy and social constraints exists that delay new electric capacity investment. In the face of investment constraints in the form of social opposition or monopolies, exporting regions are likely to face persistently higher prices that result in the transfer of economic rents from consumers to producers (Finon and Romano, 2009; Billete de Villemeur and Pineau, 2012). In the case of states that export electricity to California, regulators can consider a variety of tax- or rate-based redistribution policies to mitigate harm to consumers after western market integration.

The total effect on emissions from increased trade depends on the relative emissions contents of the affected generators in each region. If we denote $e$ as the average emissions rate from the affected generators in each region, then the change in total emissions ($\Delta E$) is represented in equation (5), where $q_i$ represents the quantity supplied before lowering barriers to trade, and $q_i^*$ denotes quantity supplied after:

$$\Delta E = e_1(q_1^* - q_1) + e_2(q_2^* - q_2) \qquad (5)$$

In the example presented in Figure 1, more electricity is generated in region 2 to sell to region 1 after reducing barriers to trade. The overall effect of increasing trade on emissions in this example is positive if $e_2 > e_1$ and negative if $e_2 < e_1$. More generally, the total emissions effect depends on the average emissions content of the marginal generators that adjust their production in response to the reduced trade barrier.

## 4. Analysis

Electricity market data covering the western U.S. during the years of 2015-2018 were collected for this analysis. Generation and price data are available for CAISO, but not for other non-CAISO balancing authorities in California, including those serving the cities of Sacramento and Los Angeles. As a result, the analytic results for prices and generation are representative of CAISO only. Imports in these models come from neighboring states as well as from balancing authorities in California outside of CAISO. Conversely, emissions data is available for all of California. In this case, the model estimates the relationship between imports and emissions for California, inclusive of all balancing authorities in the state. Furthermore, the California summary statistics presented in this section include balancing authorities in the state that are not in CAISO.

The data collected includes datasets that provide 5-minute observations of total CAISO generation by fuel type, demand, and average system price (CAISO, 2018; LCG Consulting, 2018). Table 1 shows that in CAISO, electricity supply from solar and hydro have increased while natural gas decreased over the past four years. Other fuels have remained relatively constant, including imports, which supply slightly less than 1/3 of CAISO's electricity demand. Figure 2 plots the average daily fuel mix by hour in CAISO during 2018, representing a "typical" day. It shows a daily reduction in natural gas and electricity imports during the morning when large amounts of solar come online, followed by significant increases at night when solar goes offline. If recent trends continue and solar capacity continues to displace natural gas, the need to rely on out of state electricity to balance daily changes in solar generation will grow.

The data also includes plant-level information and hourly electricity imports spanning July 2015 (the earliest this data is available) through July 2018, from the U.S. Energy Information Administration (U.S. EIA 2018 a, b). Power plants located in balancing regions that trade with California are plotted in Figure 3. All balancing authorities that trade with California are assigned to two regions, Northwest or Southwest, consistent with the organization of EIA's electricity data. Table 2 lists all the electric balancing authorities in each region that trade electricity with California, as well as each region's average net imports into California. It shows both regions have similar levels of electricity demand. Table 3 presents the capacity mix of California plus each region that trades with California from 2016, the most recent year which plant level data is available. California generates the majority of its electricity using natural gas, while neighboring regions have a more balanced electricity mix between natural gas, coal, hydro, and other fuels. Hourly environmental emissions data were collected from the U.S. Environmental Protection Agency's Air Markets Program database (U.S. EPA, 2018). I downloaded historic hourly emissions at the state level of sulfur dioxide ($SO_2$), nitrogen oxides ($NO_x$), and carbon dioxide ($CO_2$) for California and all states that trade electricity with California, from May 2014 – June 2018. Both $SO_2$ and $NO_x$ cause respiratory problems, while $CO_2$ causes climate change. All three of these pollutants are emitted from the combustion of fossil fuels, but natural gas emits only trace amounts of $SO_2$ and $NO_x$.

|  | **2014-2015** | **2015-2016** | **2016-2017** | **2017-2018** |
|---|---|---|---|---|
| **Solar** | 16,034<br>6% | 17,850<br>8% | 23,644<br>11% | 26,912<br>13% |
| **Wind** | 15,391<br>6% | 13,503<br>6% | 13,990<br>7% | 15,344<br>7% |
| **Nuclear** | 21,758<br>9% | 17,749<br>8% | 17,936<br>8% | 18,539<br>9% |
| **Hydro** | 16,004<br>6% | 17,930<br>8% | 28,453<br>13% | 25,334<br>12% |
| **Natural Gas** | 110,447<br>43% | 87,737<br>40% | 68,234<br>32% | 62,499<br>30% |
| **Imports** | 75,744<br>30% | 63,521<br>29% | 62,445<br>29% | 62,541<br>30% |
| **Total** | 255,379 | 218,290 | 214,703 | 211,168 |

*Table 1 Annual generation (GWh) and percent of total supply by fuel type, CAISO. Each column spans July 1 - June 30 of the listed years.*

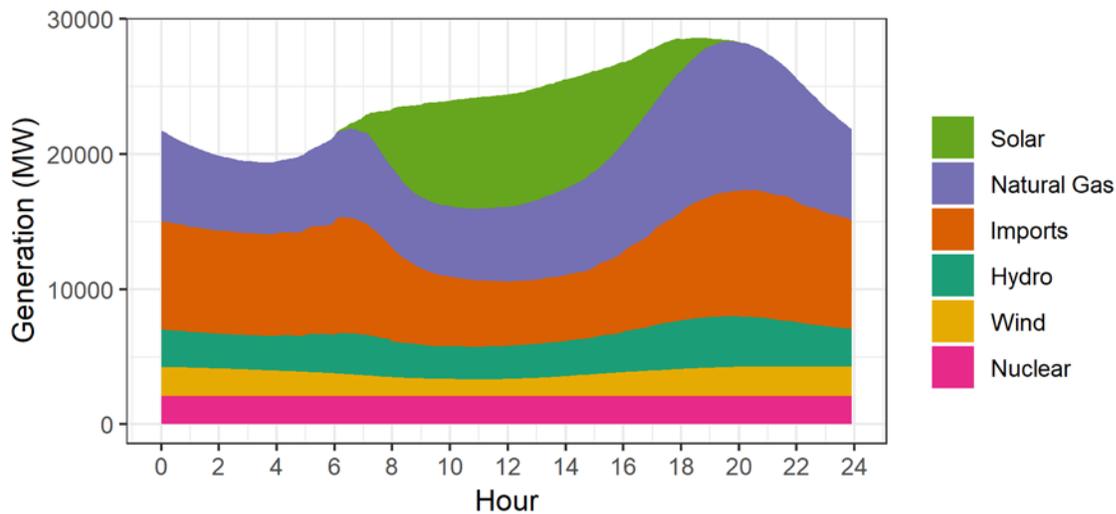

*Figure 2 Average daily generation in CAISO, 2018.*

The statistical modeling approach used for this analysis builds on a growing literature utilizing econometric-based methods with highly granular electricity market data to estimate conditional short-term relationships related to various policies and electricity prices, emissions, and generation. This includes Callaway et al., 2018; Carson and Novan, 2013; and Graff-Zivin et al., 2012. The analytic methods are discussed in the following sections.

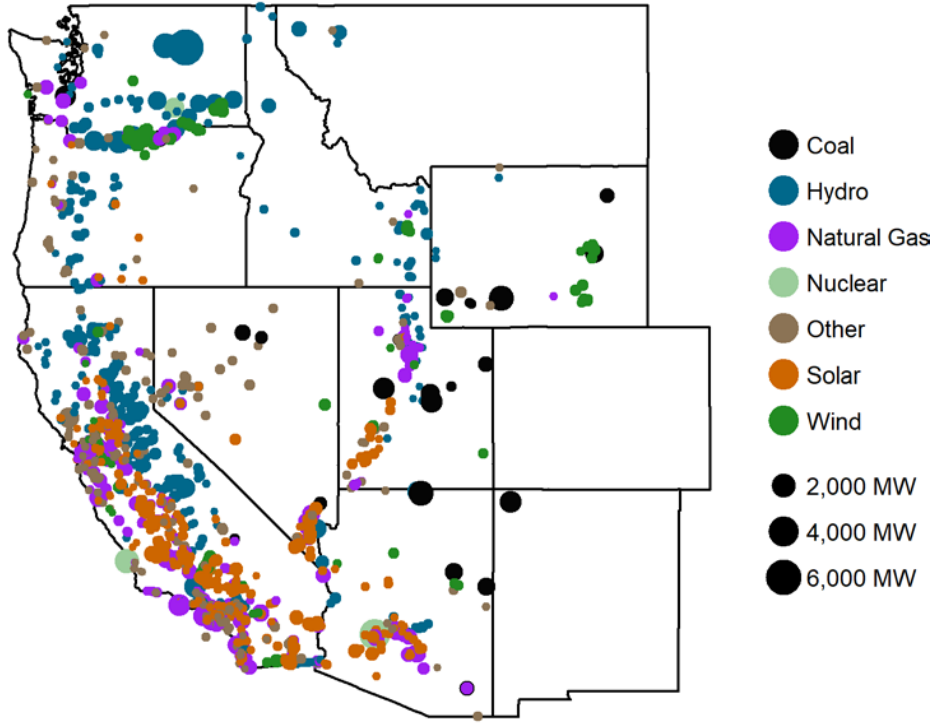

*Figure 3 Power plants in California and in balancing areas that trade with California, 2016.*

| Region | Balancing Authorities in Region | Average Net Imports (MW) |
|---|---|---|
| Northwest | Bonneville Power Administration, Nevada Power Company, PacifiCorp East, PacifiCorp West | 3,484 |
| Southwest | Arizona Public Service, Salt River Project, Western Area Power Administration - Desert Southwest | 3,205 |

*Table 2 Balancing authorities and average net imports into California by region.*

| Region | Coal | Hydro | Natural Gas | Nuclear | Other | Solar | Wind |
|---|---|---|---|---|---|---|---|
| California | 1,703 | 11,751 | 44,791 | 2,323 | 5,502 | 11,026 | 5,976 |
| | 2% | 14% | 54% | 3% | 7% | 13% | 7% |
| Northwest | 11,129 | 23,366 | 16,196 | 1,200 | 1,691 | 1,680 | 7,713 |
| | 18% | 37% | 26% | 2% | 3% | 3% | 12% |
| Southwest | 6,115 | 5,926 | 10,736 | 4,210 | 165 | 1,014 | 237 |
| | 22% | 21% | 38% | 15% | 1% | 4% | 1% |

*Table 3 Electric generating capacity (megawatts) and percent of total capacity by fuel type and region,*

## 4.1. Prices

This section describes the method for estimating the short-term relationship of increased imports on CAISO prices. The theoretical model presented in section 3 predicts that a decrease in trading costs across regions will decrease prices in the importing region, resulting in savings for consumers and revenue losses for producers. The econometric results presented in this section support this assertion. The model utilizes hourly data on imports, CAISO average system prices, and net load from July 2015 – July 2018, plotted in Figure 4. Net load is total demand minus non-dispatchable wind and solar generation. This is a more relevant variable for determining price on the supply side because it subtracts away noise in the form of wind and solar production that do not respond to short term changes in demand. Carson and Novan (2013) also utilize net load in a similar modeling framework for their study of energy storage in the Texas market.

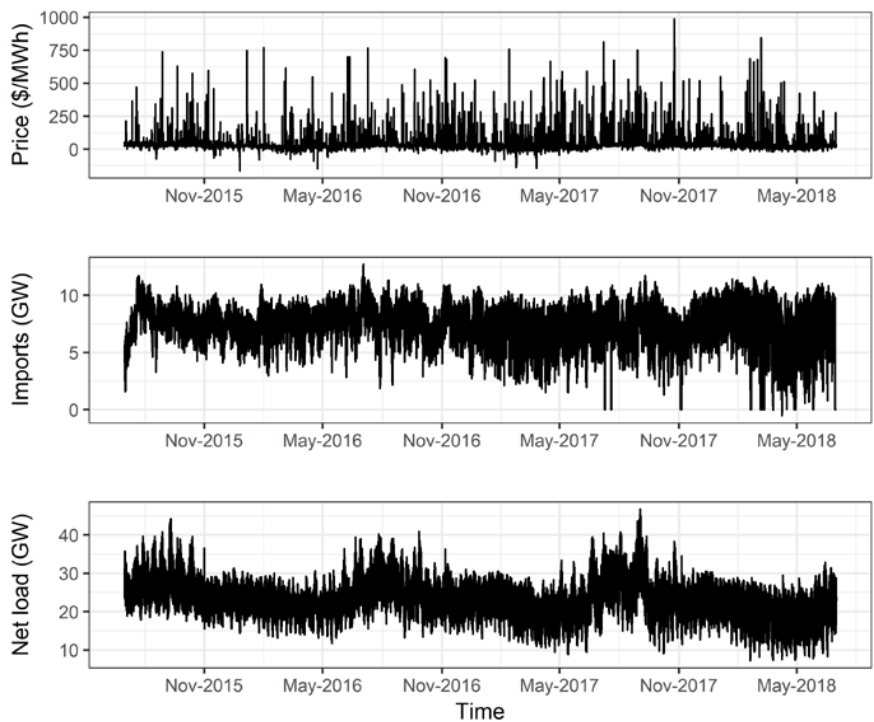

*Figure 4 Hourly CAISO average system price (top), net imports (middle), and net load (bottom).*

Electricity prices are serially correlated and have unequal variance, which will cause incorrect estimates of traditional standard errors. To obtain proper statistical inference, standard error calculation methods that are robust to heteroskedasticity and auto-correlation (HAC) are used throughout the entirety of the analysis, following the method implemented in Zeileis (2004).The data are more likely to show high levels of prices and imports during periods of high demand, confounding the bivariate relationship between price and imports. To deal with this, CAISO net load is included as a control variable. Other unobserved factors will also affect electricity price, including transmission congestion or changes in fuel prices. To account for these external factors, I include a set of date fixed effects, which difference out daily price averages from the model. Doing this accounts for price effects from a particular day, month, or year from unobserved factors like persistent congestion or changes in fuel costs. As a result, the model estimates the average within-day relationship between price and imports, conditional on hourly net load. The model specification is described in equation set (6). $\alpha_d$ represents the daily fixed effects that control for the average price each day caused by factors external to the model. The day fixed effects are programmed into the data as a set of variables equal in number to the total days in the dataset, with each variable equal to 1 during the 24 observations that occur during the respective day, and 0 otherwise.

$$price_t^* = \beta_0 + \beta_1 imports_t + \beta_2 netload_t + \alpha_d + \epsilon_t \qquad (6)$$

$$price_t^* = \ln(1 + price_t - \min(price_t))$$

Table 4 presents results from this model. Column (1) shows results from a bivariate regression model to provide intuition into the data generating process. The positive coefficient of 0.014 indicates the observed simple correlation between price and imports is actually positive. This is because high levels of prices and imports both are more likely to occur during periods of high demand, transmission congestion, higher fuel costs, and other unobserved factors that increase the cost to supply electricity. The model in column (2) controls for these effects by including net load and daily fixed effects, and shows the relationship between prices and imports conditional on these other variables is in fact negative.

For this reason, column (2) shows results from the preferred model specified in equation set (6). The coefficient on imports indicates that during the sample period from 2015-2018, a one gigawatt increase in net imports is associated with an average decrease in CAISO system price in the same hour by a multiple of $e^{0.005}$, equal to 1.005, equivalent to a 0.5% decrease. This suggests an average short-term relationship of -$0.15, or an average $4,017 in consumer savings per gigawatt-hour increase in imports. $0.15 is calculated as 0.5% of the average price observed during the data sample, $29.97/MWh. The consumer savings is calculated by multiplying the price effect by average CAISO electricity demand observed in the data sample (26,261 MW).

|                | Natural log of price |           |
|----------------|:--------------------:|:---------:|
|                | (1)                  | (2)       |
| Imports (GWh)  | 0.014*               | -0.0051*  |
|                | (0.0011)             | (0.0010)  |
| Net load (GWh) |                      | 0.015*    |
|                |                      | (0.00045) |
| Fixed Effects  |                      | Day       |
| Observations   | 26,303               | 26,303    |
| R2             | 0.032                | 0.29      |
| Adjusted R2    | 0.032                | 0.26      |

*Table notes: Heteroskedasticity and autocorrelation-robust standard errors reported in below coefficients; '*' denotes the probability of the coefficient being zero is less than 0.01.*

*Table 4 Results from price and imports models.*

These results suggest a doubling of interregional flows between CAISO and neighbors would be associated with an average CAISO price decrease of $1.09, corresponding with short-term annual consumer savings of approximately $252 million. These short-term savings are well in excess of the likely administrative costs required to setup the regional market. This is based off the $40 million one-time cost required to implement a similar market expansion in the PJM market (Mansur and White, 2012). I used a doubling of regional trade as the basis for the annual consumer savings calculation because the recent study commissioned by CAISO assumed regional market integration would roughly double the limits on interregional electricity flows (Chang et al, 2016). The immediate price reduction of $1.09/MWh from doubling regional trade is calculated by multiplying the average price marginal effect (-0.15) by the average level of net imports (approximately 7 GW) observed during 2015-2018. The annual consumer savings of $252 million is then calculated by multiplying the full price effect by average CAISO electricity demand and 8,760 hours per year. These empirically estimated consumer savings are similar in magnitude to the production cost savings predicted by the CAISO-commissioned simulation study. Unfortunately, price effects in neighboring states outside of California are not estimated in this study because public wholesale price or marginal cost data is unavailable for non-CAISO regions. The economic theory presented in section 3 predicts a price increase in these net-exporting states.

The day fixed effects parameters ($\alpha_d$ in equation 6) control for daily average changes in the outcome variable, leaving within-day variation in prices and imports to use for calculating the coefficient estimates. In this way, the model nets out all unobserved factors that confound the observed relationship between price and imports that vary on a daily level. This includes controlling for different outcomes between work days and weekends, seasonal effects, and annual macroeconomic effects. It is possible there are short-term factors not included in the model that affect both the outcome variable and imports, including within-day transmission congestion, fuel costs, and outages in California. However, theory suggests all of these factors are positively correlated with both the independent and outcome variables in that they cause higher CAISO prices and also make imports into CAISO more competitive. Thus, the existence of these factors would increase the estimated coefficient, suggesting the estimated effect provided in column (2) of Table 4 is a conservative, upper-bound estimate, and the true effect is more negative.

In general, empiric economic studies often have difficulty disentangling the relative effects of supply-side factors (like imports) from demand-side factors, because both sets of factors simultaneously interact to determine price. However, in the case of wholesale electricity markets, most electricity consumers face prices that do not track short-term changes in wholesale prices. The lack of price response on the demand side minimizes the simultaneity bias concern. If we consider a case where consumers did in fact respond to short term changes in price, theory suggests simultaneity would positively bias the model estimate relative to the true effect. This is because if consumers did respond to short-term wholesale price signals, the reduction in price from increasing imports would be mitigated by a positive demand response. In this case, the true effect would also be more negative than the estimated relationship.

Some degree of endogeneity is likely present between imports and electricity prices. In the short-term a CAISO price increase will incent additional imports into CAISO. In these models, a significant portion of electricity price variation is accounted for via the inclusion of CAISO demand as a control variable. However, unplanned generation outages and transmission congestion are examples of other factors that can cause high prices. These effects cannot be directly controlled for due to data unavailability, but they are largely controlled for in an indirect manner by the inclusion of day fixed effects. In this context, the results can be interpreted as the within-day average effect of imports plus other within-day unobserved effects on price. To the extent that within-day unobserved variables that are correlated with imports cause price increases (including generator outages and transmission congestion), the short-term relationship estimate in column 2 of Table 4 would be positively biased, and the true effect of imports would be more negative.

### 4.2. Emissions

In this part of the analysis, hourly data on carbon dioxide ($CO_2$), sulfur dioxide ($SO_2$), and nitrogen oxide ($NO_x$) emissions from electricity generation by region are utilized to estimate the relationship between electricity imports and emissions. Hourly $CO_2$ emissions in California, the northwest, and the southwest regions from July, 2015 until July, 2018 are plotted in Figure 5. Average emissions levels during the sample period for each region and pollutant are reported in Table 5.

Figure 5 shows the $SO_2$ and $NO_x$ series are highly correlated with $CO_2$ emissions and follow similar patterns. Like the price data series, the distributions of emissions are positively skewed and exhibit similar patterns of serial correlation. To deal with these issues, a log transformation of emissions and HAC robust standard errors are utilized, similar to the procedure described in section 4.1. More specifically, models following the structures described in equation set (7) are estimated.

$$\ln(em_{i,t,CA}) = \beta_0 + \beta_1 imports_{t,CA} + \beta_2 netload_{t,CA} + \alpha_d + \epsilon_t^a \quad (7)$$

$$\ln(em_{i,t,r}) = \delta_0 + \delta_1 exports_{t,r} + \alpha_d + \gamma_h + \epsilon_t^b$$

$$i = \{CO_2, SO_2, NO_x\}, \quad r = \{NW, SW\}, \quad d = \{Jul\ 1, 2015 : Jun\ 30, 2018\}, \quad h = \{1:24\}$$

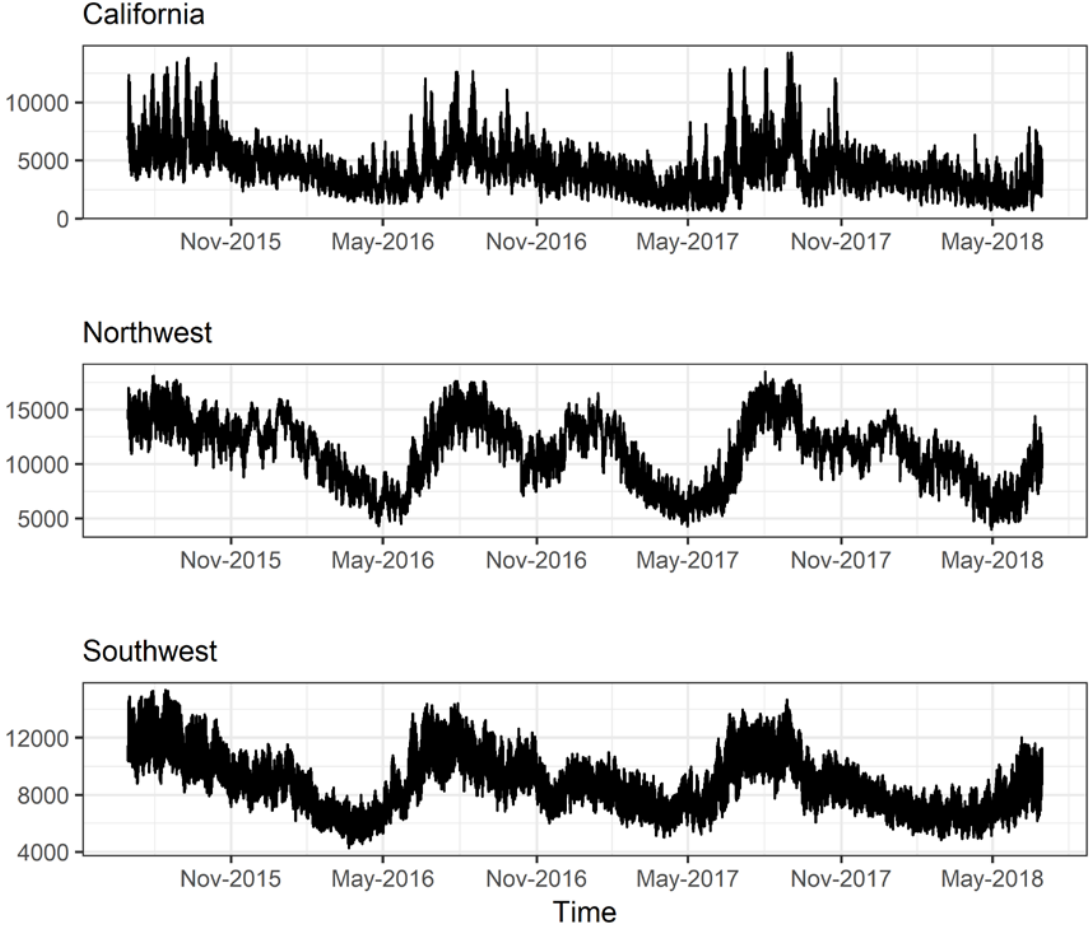

*Figure 5 Hourly CO2 emissions by region, metric tons.*

|  | CO$_2$ | SO$_2$ | NO$_x$ |
|---|---|---|---|
| California | 4,018 | 42 | 622 |
| Northwest | 11,138 | 11,281 | 14,279 |
| Southwest | 8,751 | 5,248 | 16,022 |

*Table 5: Average hourly emissions by pollutant and region, 2015 – 2018, CO$_2$ emissions are reported in metric tons, SO$_2$ and NO$_x$ are reported in pounds.*

In the first line of equation set (7), $em_{i,t,CA}$ represents hourly emissions in California, where $i$ indexes each pollutant. $imports_{t,CA}$ represents hourly total net imports into California, $netload_{t,CA}$ is CAISO's hourly net load, and $\alpha_d$ is a set of day fixed effects, one for each day in the data sample. In the second line, $em_{i,t,r}$ represents hourly emissions by region, with $r$ indexing the Northwest and Southwest regions. $exports_{t,r}$ represents hourly exports from region $r$ into California. Hourly net load data for the northwest and southwest regions are not publicly available. To make up for this, a set of 24 hour fixed effects are included to control for average intra-day variation in demand. For each region, the models are simultaneously solved for the three pollutants as a set of seemingly unrelated regressions utilizing the method described in by Henningsen and Hamann (2007), and the associated software they built. The seemingly unrelated regression approach yields more precise estimates compared to a set of independent regressions by modeling the covariance between pollutants.

Table 6 presents results for each region and pollutant. Columns 2, 4, and 6 include the preferred model specifications for $CO_2$, $SO_2$, and $NO_x$ emissions, respectively. The results show a significant decrease in California emissions associated with electricity imports. Conversely, the northwest and southwest regions show a significant increase in emissions associated with exports. These estimates suggest that, on average, electricity trade into California is being supplied by a nonzero portion of fossil generation in exporting regions that displaces some fossil generation within California. Each coefficient $\beta$ can be interpreted after an exponential transformation ($e^\beta$) as the average multiplicative increase in price associated with a one-gigawatt increase in imports. These are most easily understood as percentage changes. Considering column 2 for example, a one gigawatt increase in imports into California is associated with an 8.3% ($e^{0.080} = 1.083$) decrease in $CO_2$ emissions in California, a 2.6% increase in $CO_2$ emissions in the northwest, and a 2.4% increase in $CO_2$ emissions in the southwest. Multiplying these percentage changes by the average hourly $CO_2$ emissions level from 2015-2018 (previously displayed in Table 5) indicates that, on average, a 1 gigawatt-hour (GWh) increase in net imports into California is associated with a 321 metric ton reduction of California $CO_2$ emissions . This is close to the $CO_2$ emissions rate for the average combined cycle gas plant in the U.S. (U.S. Department of Energy, 2016). Thus, it is likely that electricity imports are displacing marginal generation from combined cycle gas plants in California.

All the estimated emissions effects for each pollutant and region are presented in Table 7. The decrease in California $CO_2$ is partially offset by emissions increases in its neighboring regions. One GWh of exports to California is associated with a 284 metric ton increase in the Northwest region, or a 214 metric ton increase in the Southwest. A direct comparison of emissions effects between California and its neighbors requires taking the average of the emissions changes for the exporting regions, weighted by average California trade levels, shown in the fourth row of Table 7. Doing this suggests that every 1 GWh increase in trade is associated with a net reduction in $CO_2$ emissions by 70 tons, and net increases in $SO_2$ and $NO_x$ emissions of 283 and 270 lbs., respectively. The estimated effects for each pollutant and region are presented in Table 7, with the overall net changes for each pollutant calculated in the bottom row.

The positive relationship between trade and $SO_2$ and $NO_x$ emissions provide evidence that some coal plants in both the Northwest and Southwest regions are increasing on the margin when exports to California increase. This is because natural gas plants only emit trace amounts of these pollutants. Coal plants range widely in $SO_2$ and $NO_x$ emissions rates, depending on the environmental technology at the plant and type of coal combusted. In 2015, the average $SO_2$ emissions rate for coal in the U.S. was approximately 3,622 lbs./GWh (U.S. EIA, 2017). Using this national average as an estimate of the rate in the northwest and southwest regions suggests that less than 10% of each GWh of California imports on average is supplied by coal.

| California | Natural log of $CO_2$ emissions | | Natural log of $SO_2$ emissions | | Natural log of $NO_x$ emissions | |
| --- | --- | --- | --- | --- | --- | --- |
| | (1) | (2) | (3) | (4) | (5) | (6) |
| Imports (GWh) | 0.030 | -0.080* | 0.024 | -0.078* | 0.017 | -0.15* |
| | (0.013) | (0.0030) | (0.012) | (0.0029) | (0.019) | (0.0044) |
| Net load (GWh) | | 0.071* | | 0.070* | | 0.075* |
| | | (0.0012) | | (0.0012) | | (0.0015) |
| Fixed Effects | | Day | | Day | | Day |
| $R^2$ | 0.017 | 0.94 | 0.011 | 0.94 | 0.0038 | 0.79 |

| Northwest | Natural log of $CO_2$ emissions | | Natural log of $SO_2$ emissions | | Natural log of $NO_x$ emissions | |
| --- | --- | --- | --- | --- | --- | --- |
| Exports (GWh) | -0.057* | 0.026* | -0.062* | 0.034* | -0.066* | 0.030* |
| | (0.017) | (0.0026) | (0.019) | (0.0027) | (0.019) | (0.0027) |
| Fixed Effects | | D,H | | D,H | | D,H |
| $R^2$ | 0.069 | 0.96 | 0.058 | 0.95 | 0.071 | 0.95 |

| Southwest | Natural log of $CO_2$ emissions | | Natural log of $SO_2$ emissions | | Natural log of $NO_x$ emissions | |
| --- | --- | --- | --- | --- | --- | --- |
| Exports (GWh) | 0.074* | 0.024* | 0.11* | 0.035* | 0.095* | 0.018* |
| | (0.012) | (0.0049) | (0.015) | (0.0072) | (0.015) | (0.0066) |
| Fixed Effects | | D,H | | D,H | | D,H |
| $R^2$ | 0.17 | 0.92 | 0.19 | 0.84 | 0.14 | 0.94 |

*Table notes:*

*Heteroskedasticity and autocorrelation robust standard errors reported in parenthesis.*

*\* denotes the probability of the coefficient being zero is less than 0.01*

*"D,H" stands for day and hour fixed effects.*

*All models are estimated with 26,253 data observations.*

*Adjusted $R^2$ values are within 0.01 of the reported simple $R^2$ for all models.*

*Table 6 Results from emissions models.*

|  | $CO_2$ | $SO_2$ | $NO_x$ |
|---|---|---|---|
| California | -321 | -3 | -90 |
| Northwest | 284 | 383 | 421 |
| Southwest | 214 | 181 | 294 |
| Weighted Avg - NW & SW | 251 | 286 | 360 |
| **Net change (row 1 plus row 4)** | **-70** | **283** | **270** |

*Table 7 Estimated change in emissions due to 1 GWh increase in trade. CO2 is measured in metric tons, SO2 and NOx are measured in pounds.*

$SO_2$ emissions are subject to national caps in the United States under the acid rain program. As a result, increasing regional trade between U.S. states will not lead to long-term changes in these emissions. Instead, the short-term increases in $SO_2$ associated with increasing regional trade must be offset by emissions reductions elsewhere in order to keep pollutant levels under the cap. As regional trade increases, emitting producers will increase profits by selling at a higher price to California consumers. These profits will be offset somewhat by having to pay for emissions reductions elsewhere in order to meet the $SO_2$ cap. $NO_x$ emissions are not subject to a national or regional cap in the western U.S. As a result, increases in $NO_x$ emissions due to regional trade are more likely to be sustained long term. To eliminate long-term $NO_x$ emissions increases from regional electricity trade, it is important that an effective $NO_x$ emissions cap is put in place throughout the regional market.

California currently caps domestic $CO_2$ emissions as well as $CO_2$ emissions from out of state producers who sell into California. Neighboring states do not have caps in place (Fowlie and Cullenward, 2018). Despite the lack of $CO_2$ policy in neighboring states, the fact that measured $CO_2$ emissions impacts from increased regional trade are still net negative suggests that California's cap and trade program has been relatively effective in limiting the carbon content of imported electricity, and minimizing emissions leakage to neighbors. Despite this evidence suggesting minimal leakage, recent research suggests leakage may be an important issue for California (Hogan, 2017; Tarufelli and Gilbert, 2017).

In Table 6, columns 1, 3, and 5 report results from simple bivariate regressions of emissions, to provide additional intuition into the data generating processes. In California and the Southwest, results from the bivariate regressions are greater than the multiple regressions. This is likely due to similar reasons as the price model in section 4.1: periods with both high emissions and high imports are positively correlated with periods of high demand and other supply factors that increase cost, which positively bias the bivariate results. Once the models condition on these other variables, the positive inflationary effect disappears. The northwest region shows the opposite effect in that the bivariate regression result is less than the multiple regression result. Unlike in California and the southwest, the northwest region has peak electricity demand during the winter due to electric heating. Figure 6 plots relative monthly demand levels for these regions. It shows the northwest region demand peaks in the winter while the other regions peak in the summer. As a result, periods with high exports into California occur during periods with relatively lower local emissions in the northwest, resulting in an opposite, deflationary effect impacting the bivariate model relative to the multiple regression model.

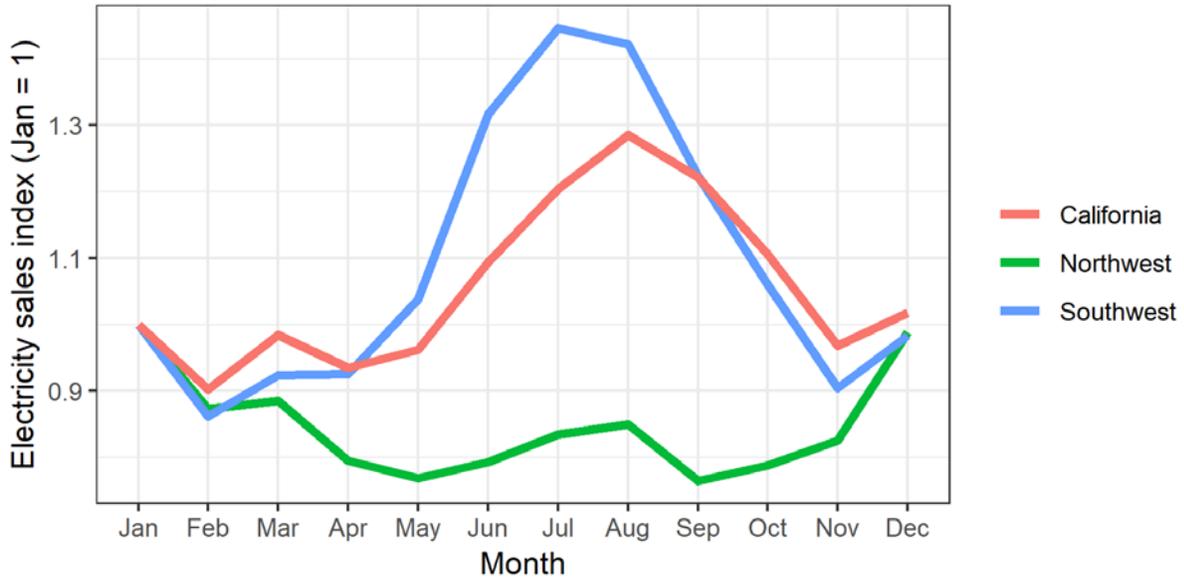

*Figure 6 Index of average monthly electricity sales by region, 2015-2017.*

Examining the residuals of the regression models illustrates the benefit of utilizing day fixed effects. The top panel of Figure 7 plots the residuals from a regression model of $CO_2$ emissions with imports and net load as covariates, while the bottom plots the residuals from the same model except day fixed effects are included. The residuals in the top panel show non-stationary trends, in that different subsets of the data have non-zero means. This is problematic for model estimation. The residuals from the model with day fixed effects show a stationary series that more closely approximates white noise, indicating more efficient model estimates. The residuals still exhibit heteroskedasticity in that the variance of the series is not constant, and autocorrelation in that values are correlated with prior values. These issues are present across all the models estimated in this analysis, and are addressed by using HAC robust standard errors for inference of coefficient estimates.

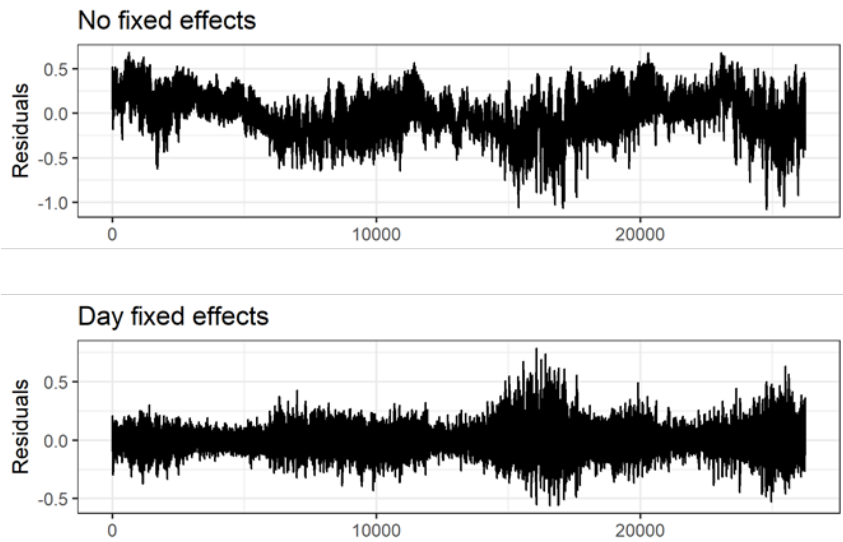

*Figure 7 Residuals from California CO2 models with and without fixed effects.*

## 4.3. Generation

The set of generation models for this analysis are designed to better understand the relationship between regional electricity trade and dispatchable electric generation in CAISO. Hourly generation data for nuclear, hydro, and natural gas generation are utilized, and plotted in Figure 8. The same electric interchange data from EIA, along with hourly generation data from CAISO, are used. The model is summarized in equation (8).

$$gen_{i,t} = \beta_0 + \beta_1 imports_t + \beta_2 netload_t + \alpha_d + \epsilon_{i,t} \qquad (8)$$

$$i = \{nuclear, hydro, natural\ gas\}, \quad d = \{Jul\ 1, 2015 : Jun\ 30, 2018\}$$

The three equations for each type of generation are simultaneously estimated as a set of seemingly unrelated regressions, the results of which are presented in Table 8. Like in previous sections, results from bivariate regressions are also included, although the models including net load day fixed effects presented in columns 2, 4, and 6 represent the preferred specifications. For all three fuel types, the bivariate model results are larger than the models with additional control variables. This is due to the inflationary effect from the fact that high levels of both imports and generation occur during periods of high demand.

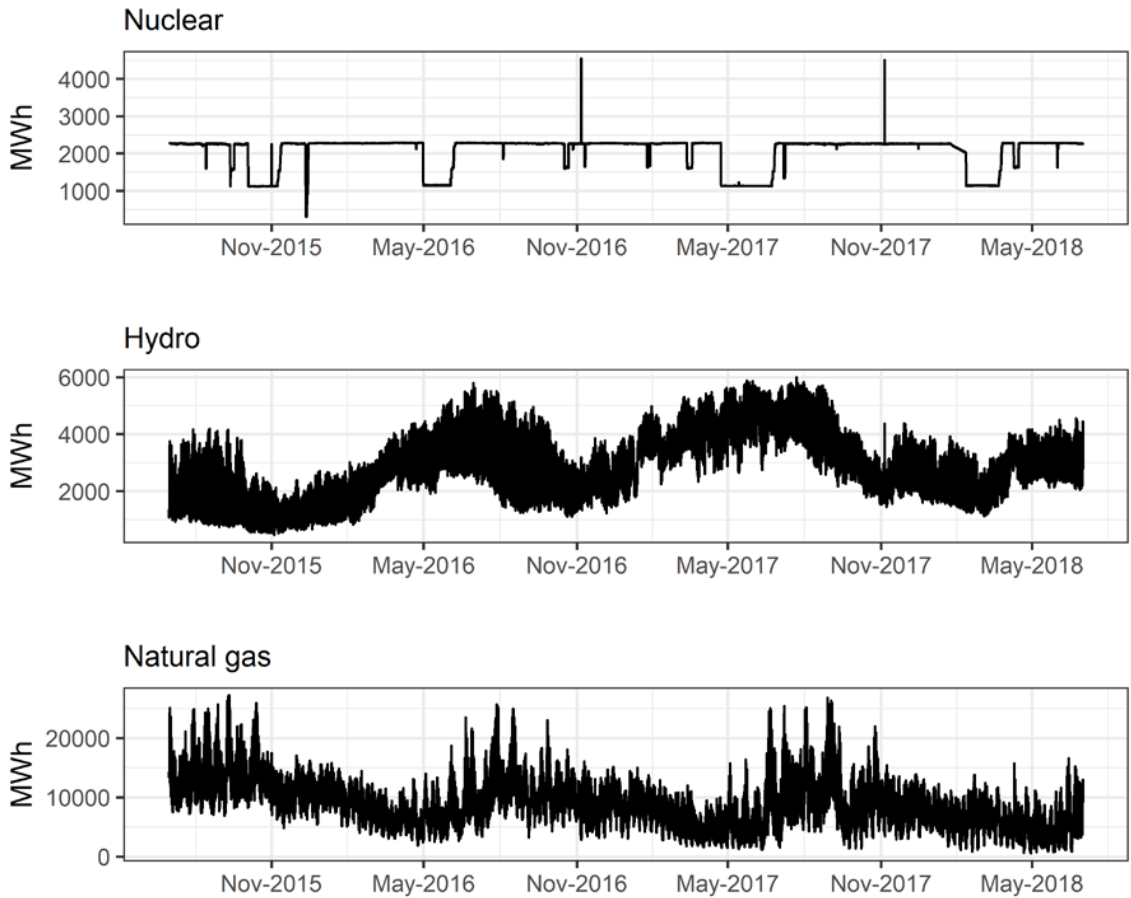

*Figure 8 Hourly CAISO generation by fuel type.*

|  | **Nuke (GWh)** | | **Hydro (GWh)** | | **Gas (GWh)** | |
| --- | --- | --- | --- | --- | --- | --- |
|  | (1) | (2) | (3) | (4) | (5) | (6) |
| Imports (GWh) | 0.00 | 0.00 | 0.072* | -0.077* | 0.47* | -0.61* |
|  | (0.01) | (0.00) | (0.023) | (0.0039) | (0.11) | (0.011) |
| Netload (GWh) |  | 0.00 |  | 0.15* |  | 0.70* |
|  |  | (0.00) |  | (0.0021) |  | (0.0054) |
| Fixed Effects |  | Day |  | Day |  | Day |
| R2 | 0.00 | 0.99 | 0.014 | 0.98 | 0.045 | 0.96 |

*Table Notes:*

*Heteroskedasticity and autocorrelation robust standard errors reported in parenthesis.*
*\* denotes the probability of the coefficient being zero is less than 0.01.*
*All models are estimated with 26,300 data observations.*
*Adjusted R2 values are within 0.01 of the simple R2 for all models.*

*Table 8 Results from generation models.*

The results in Table 8 show that electricity imports have no observed short-term relationship with nuclear energy. As shown in the first panel in Figure 8, nuclear energy in CAISO often remains constant, and is not subjected to intra-day fluctuations. Occasionally, nuclear shows large changes in output, driven by a relatively few large units turning on and off. These changes occur too infrequently for any meaningful short-term statistical relationship to be estimated. As a result, the model returns a result of zero. The remaining results for hydro and natural gas suggest that every GWh of electricity imports is associated with an average 0.69 GW decrease in dispatchable generation in CAISO. Approximately 0.08 GW of this decrease is from hydro and the remaining 0.61 GW is from natural gas. The fact that natural gas makes up the majority of generation displaced by imports is consistent with the emissions results estimated in section 4.2.

## 5. Conclusions and Policy Implications

In summary, this paper analyzes short-term market relationships relevant to increasing regional electricity trade between California and neighboring states. Specifically, it provides evidence characterizing potential short-term effects of increased regional trade on prices, emissions and generation. The study finds that from 2015 – 2018, a one GWh increase in California imports was associated with an average $0.15/MWh decrease in the CAISO system electricity price, or $4,017 in consumer savings. Extrapolating these results suggest that a doubling of imports would produce approximately $252 million in annual savings for CAISO consumers. This estimate does not include long-term effects that would accrue from changes in investment decisions due to changing regional trade patterns, which other studies suggest will offset price effects in the long-term while produce additional avenues for savings for California consumers by enabling more cost-effective capacity investments. Due to data limitations, this study does not consider price impacts outside of California from increased regional trade. Electricity market integration studies from other regions, along with economic theory and the fact that California is a net importer of electricity on average suggests that increased regional trade will cause higher prices outside of

California. This will partially offsetting the savings experienced in California and generate political economy concerns related to short-term rent transfers from consumers to producers outside of California.

This analysis also finds that a 1 GWh increase in trade is associated with a 321 metric ton reduction in $CO_2$ emissions from California power plants. Taking account of the offsetting effect from increased $CO_2$ emissions in neighboring regions suggests a net 70 ton decrease in CO2 emissions for each GWh increase in regional trade. Short-term net increases in $NO_x$ and $SO_2$ outside of California are also observed, suggesting a small portion of exports to California is supplied by coal generation. As a result, increasing trade through a regional market will likely increase long term $NO_x$ emissions absent a $NO_x$ emissions cap.

From the perspective of a researcher or analyst, one way centralized electricity markets are useful is that they produce lots of highly granular data that provide the basis for studies like this. It is currently difficult to estimate effects in non-market regions outside of California because public data is scarce. Regulatory bodies like the Federal Energy Regulatory Commission and state public utility commissions should work to increase the availability of market data to enable more informed policy decisions. A possible next step after this analysis includes a more detailed empirical examination of electric producers trading with California. As the state continues trading electricity with its neighbors and continues its ambitious emissions reductions goals, it is important to better characterize generator responses to California electricity policies outside of California. This will lead to a better understanding of the full regional impacts from California's evolving and dynamic energy policies.

The empiric results of this study suggest significant savings for consumers can be achieved through regional electricity market integration, likely well in excess of market implementation costs. However, due to data limitations this analysis was not able to estimate consumer costs of regional trade outside of California, nor increases in profits to producers who can sell electricity at higher prices in California. This analysis provides empirical evidence suggesting improving electricity trade across the western U.S. through a regional market will lead to significant near-term monetary benefits, and help reduce $CO_2$ emissions across the region. It concludes that efforts to expand California's market to the western U.S. should move forward in parallel with strong emissions policies that cover the full market region.

## 6. Acknowledgements

The author thanks Eric Gimon of Energy Innovation for his thoughtful feedback on this paper and several anonymous referees for their review.